<div class="moz-text-flowed" style="font-family: -moz-fixed">
\documentclass[reqno,11pt,centertags]{amsart}

\usepackage{amssymb}

\def\R{I\kern-.30em{R}}
\def\N{I\kern-.30em{N}}
\def\F{I\kern-.30em{F}}
\def\C{I\kern-.60em{C}}
\def\Z{Z\kern-.50em{Z}}
\def\P{I\kern-.30em{P}}
\def\E{I\kern-.30em{E}}
\def\build#1_#2^#3{\mathrel{\mathop{\kern
0pt#1}\limits_{#2}^{#3}}}
\def\Sum{\displaystyle\sum}

\newcommand{\Schr}{Schr\"{o}dinger}

\newcommand{\ea}{\end{array}}
\newcommand{\bea}{\begin{eqnarray}}
\newcommand{\eea}{\end{eqnarray}}

\newcommand{\beq}{\begin{equation}}
\newcommand{\eeq}{\end{equation}}

\renewcommand{\Im}{\text{\rm Im}}

\numberwithin{equation}{section}

\theoremstyle{definition}

\theoremstyle{remark}

\newcommand{\abs}[1]{\lvert#1\rvert}

\begin{document}

\begin{titlepage}

\begin{center}

  {\bf Continuity with respect to Disorder \\
  of the Integrated Density of States }

  \vspace{0.3 cm}

  \setcounter{footnote}{0}
  \renewcommand{\thefootnote}{\arabic{footnote}}


  {\bf Peter D.\ Hislop \footnote{Supported in part by NSF grant
      DMS-0202656.}}

  \vspace{0.1 cm}

  {Department of Mathematics \\
    University of Kentucky \\
    Lexington, KY 40506--0027 \\
   USA}

\vspace{0.1 cm}

{\bf Fr\'ed\'eric Klopp } 

  \vspace{0.1 cm}

  {D\'epartement de Math\'ematiques   \\
   Universit\'e Paris XIII  \\
  Institut Galil\'ee\\
   F-93430 Villetaneuse \\
France}

\vspace{.1in}

{\bf Jeffrey H. Schenker }  

  \vspace{0.1 cm}

  {Theoretische Physik \\
    ETH Z\"urich \\
    CH-8093 Z\"urich \\
    Switzerland }

\end{center}

\vspace{0.2 cm}
\begin{center}
  {\bf Abstract}
\end{center}

\noindent We prove that the integrated density of states (IDS) associated to 
a
random Schr\"odinger operator is locally uniformly
H\"older continuous as a function of the
disorder parameter $\lambda$. In particular, we obtain convergence of the
IDS, as $\lambda
\rightarrow 0$, to the IDS for the unperturbed operator at all energies for
which the IDS for the unperturbed operator is continuous in energy.

\vspace{0.5 cm}

\noindent \today

\end{titlepage}

\section{Introduction and Results.}

In this letter, we use the methods recently developed in \cite{[CHK],[CHN]} 
to
prove that the integrated density of states (IDS) $N_\lambda(E)$ for a 
random
Schr\"odinger operator $H_\omega (\lambda) = H_0 + \lambda V_\omega$ is a
uniformly H\"older continuous function of the disorder parameter $\lambda$
at energies $E$ for which the unperturbed operator $H_0$
has a continuous IDS $N_0(E)$, under fairly general conditions.
Moreover, the uniformity in $\lambda$ implies
that $N_\lambda (E) - N_0 (E)$ is (H\"older) continuous
in $\lambda$, as $\lambda \rightarrow 0$, at points $E$ of (H\"older)
continuity of $N_0(E)$.
This result applies to random Schr\"odinger
operators on the lattice $\Z^d$ and on the continuum $\R^d$, given as
perturbations of a deterministic, background operator $H_0 = ( - i \nabla - 
A_0
)^2 + V_0$. We assume that the background operator is self-adjoint with
operator core $C_0^\infty ( X )$ (smooth, compactly supported functions
on $X$) for $X = \Z^d$ or $X= \R^d$. For simplicity, we assume
that $H_0 \geq  - M_0 > - \infty$,
for some finite constant $M_0$. In addition, we require that
$H_0$ is gauge invariant under translations by elements of $\Z^d$.
Specifically, this means that for every $m \in \Z^d$, we have $V_0(x + m) =
V_0(x)$ and $A_0(x + m) = A_0(x) +\nabla \phi_m(x)$ for some function 
$\phi_m$.
For $X=\Z^d$, the operator
$(-i \nabla - A_0)^2$ represents a short-range, e.g. nearest
neighbor, hopping matrix.

We consider Anderson-type random potentials $V_\omega$ constructed from a
family of independent, identically distributed (iid) random variables $\{
\omega_j ~| ~ j \in \Z^d \}$. On the lattice $\Z^d$, the potential acts as
\beq
\label{potlattice1}
(V_\omega f)(m) = \omega_m f(m), ~~m \in \Z^d,  ~~f \in
\ell^2 (\Z^d),
\eeq
On $\R^d$, the potential $V_\omega(x)$ also depends
on the single-site potential $u$, and is also a multiplication
operator given by
\begin{equation}
\label{potcont1}
( V_\omega f) (x) = \Sum_{j \in \Z^d  } \; \omega_j  u ( x - j ) f(x),
~~f \in L^2 ( \R^d).
\end{equation}
Precise hypotheses on the single-site potential $u$ and the random variables
$\{ \omega_j ~| ~ j \in \Z^d \}$ are given below.

The family of random \Schr\ operators is given by
\begin{equation}
\label{schrop1} H_\omega ( \lambda ) = H_0 + \lambda V_\omega .
\end{equation}
The parameter $\lambda > 0$ is a measure of the disorder strength, and we
consider the other parameters entering into the construction of $V_\omega$,
that is $\| u \|_\infty$ and the distribution of $\omega_0$, as fixed. As we
are interested in the explicit dependence on $\lambda$, we will write
$H_\lambda$ for $H_\omega (\lambda)$ and suppress $\omega$ in the notation. 
Due
to the assumed gauge invariance under shifts of $H_0$, and the explicit
form of the random potential given in (\ref{potlattice1}) and
(\ref{potcont1}), the random operator
$H_\omega(\lambda)$, for fixed $\lambda$, is ergodic with respect to the 
gauge
twisted shifts \beq \psi(x) \mapsto e^{i \phi_m(x)}\psi(x - m) \; , \quad 
\psi
\in L^2(X) . \eeq

We mention that the results of this note are easily modified to apply to the
random operators describing acoustic and electromagnetic waves in disordered
media, and we refer the reader to \cite{[CHT],[FK1],[FK2],[GK]}.

Our result follows the investigation initiated in \cite{[CHK]} where a proof 
of
the H\"older continuity in energy of the IDS is given that relies on the
continuity of the IDS for the unperturbed, background operator $H_0$. As in 
the
first part of \cite{[CHK]}, we require that the IDS $N_0(E)$ for the 
background
operator $H_0$ exists and that it is H\"older continuous in the energy. The
proof is local in the energy and applies at any energy $E$ at which $N_0 
(E)$
is H\"older continuous. In particular, it applies to Landau Hamiltonians 
away
from the Landau levels, where $N_0(E)$ is discontinuous.

Before stating our results, let us make precise the hypotheses on the random
potential.

\vspace{.1in}

\noindent {\bf Hypothesis (H1).} {\it The family of iid random variables $\{
\omega_j ~| ~j \in \Z^d \}$ is distributed with a density $h \in L^\infty (
\R )$ with compact support.}

\vspace{.1in} \noindent {\bf Hypothesis (H2).} {\it The single-site 
potential
$u \geq 0$ is bounded with compact support. There exists an open subset
$\mathcal{O} \subset ~\mbox{supp} ~u$ and a positive constant $\kappa > 0 $ 
so
that $u_{| \mathcal{O}} > \kappa > 0$. }

\vspace{.1in} \noindent We first recall a result of \cite{[CHK]}, that 
H\"older
continuity in energy of the IDS for $H_0$ implies continuity of the IDS for
$H_\omega (\lambda)$, with a constant and H\"older exponent independent of
$\lambda$.

\vspace{.1in}
\noindent
{\bf Theorem 1.1.} {\it We assume that the
Schr\"odinger operator $H_0$ admits an IDS $N_0(E)$ that is H\"older 
continuous
on the interval $I \subset \R$ with H\"older exponent $0 < q_1 \leq 1$, that 
is
\beq
\label{cont1}
|N_0(E) - N_0(E') | \leq C_0 (q_1,I) |E-E'|^{q_1},
\eeq
for all
$E, E' \in I$, and some finite constant $0 < C_0(q_1,I) < \infty$. We assume
hypotheses (H1) and (H2) on the random potential $V_\omega$. Then, for
any constant $0 < q \leq q_1 q^* / (q_1+ 2)$, where $q^* = 1$ for
$\ell^2 (\Z^d)$ and $0 < q^* < 1$ for $L^2 (\R^d)$
(see (\ref{unperttrace9})),
there exists
a finite positive constant $C(q,I)$, independent of $\lambda$, so that for 
any
$\lambda \neq 0$, and any $E, E' \in I$, we have
\beq
\label{chk1}
| N_\lambda (E) - N_\lambda (E') | \leq C(q,I) |E-E'|^q  .
\eeq
}

\vspace{.1in}
\noindent
Note that the exponent $q$ obtained by this method is roughly $1/3$ whereas
it is believed that it should hold with $q=1$ (see section 4).
A similar result was obtained recently by one of us
\cite{[S]}, using a method quite different from that in \cite{[CHK]}.

We now present the main result of this note.

\vspace{.1in}
\noindent
{\bf Theorem 1.2.} {\it Under the hypotheses of Theorem
1.1, for any bounded interval $J \subset \R$, there exists a finite, 
positive
constant $C(q,I,J)$, such that if $\lambda, \lambda' \in J$, we have \beq
\label{disordercont11} | N_\lambda (E) - N_{\lambda'} (E) | \leq C(q,I,J) |
\lambda - \lambda'|^{q_2} , \eeq for any $E \in I$ and $0 < q_2 \le 2 q /
(q+3)$, where $0 < q \leq q_1 q^* / (q_1 +2) $
is the exponent in (\ref{chk1}). }

Until recently, it was not known that the IDS remained bounded in the weak
disorder limit $\lambda \rightarrow 0$. In particular, result
(\ref{disordercont11}) was known only for closed intervals $J$ {\it disjoint
from zero}. This result follows from the Helffer-Sj\"ostrand formula (see
section 3 and also \cite{[GK]}). However, the constant $C(q_1,I,J)$ obtained
from that proof scales like $[ \mbox{dist} ~(J, 0)]^{-1}$.

Recall that control of the IDS comes from the Wegner estimate, \beq \P \{
\mbox{dist} ~ ( \sigma ( H_\Lambda) , E ) \leq \eta \} \leq C_q (\lambda) |
\Lambda  | \eta^q , \eeq for any $0 < q \leq 1$.  Here $H_\Lambda$ is the
restriction, with suitable boundary conditions, of $H_\omega(\lambda)$ to a
bounded open set $\Lambda$ of volume $|\Lambda|$. In the usual proof of the
Wegner estimate \cite{[CHN],[Wegner]}, the constant $C_q (\lambda)$ diverges 
as
$1 / \lambda$ as $\lambda \rightarrow 0$. In \cite{[CHK]}, a different proof 
of
the Wegner estimate is given for which the constant $C(q,I)$ is {\it 
uniformly
bounded in} $\lambda$. The only deficit of this proof is that the H\"older
exponent $q$ for the IDS $N_\lambda (E)$
must be taken sufficiently small (as stated in Theorem 1.1)
relative to the assumed H\"older exponent $0 < q_1 \leq 1$ of the IDS of
$N_0(E)$ in (\ref{cont1}).  In particular, the bound gives no
information about the density of states (DOS) $\rho_\lambda (E)
\equiv d N_\lambda (E)/d E$ (see section 4 for a
further discussion of the DOS).

We have the following two corollaries of Theorems 1.1 and 1.2.

\vspace{.1in}

\noindent
{\bf Corollary 1.3.} {\it Under the same assumptions as
Theorem 1.1, let $J \subset \R$ be any closed, bounded interval containing 
$0$.
Then, there exists a finite, positive constant $C(q,I,J)$, so that we have 
for
any $E \in I$ and $\lambda \in J$,
\begin{equation}
\label{disordercont1} | N_\lambda (E ) - N_0(E) |  \leq  C(q_1,I,J) | 
\lambda
|^{q_2 },
\end{equation}
where $0<q_2 \le 2q / (q+3)$, where $0 < q \le q_1 q^* / (q_1+2)$
is the exponent
in (\ref{chk1}). }

\vspace{.1in}

\noindent
There is a version of Theorem 1.2 and Corollary 1.3 with the weaker
hypothesis of continuity for $N_0(E)$ and
with a correspondingly weaker result.

\vspace{.1in}

\noindent
{\bf Corollary 1.4.} {\it We assume that the
Schr\"odinger operator $H_0$ admits an IDS $N_0(E)$ that is continuous at 
$E$.
Then, under the same hypotheses (H1) and (H2) as in Theorem 1.1, we have for
any $\lambda$ that the IDS $N_\lambda$ is also continuous at $E$ and that 
\beq
\label{wdisordercont1} \lim_{\lambda' \rightarrow \lambda } N_{\lambda'} (E) 
=
N_\lambda  (E). \eeq }

\vspace{.1in} \noindent In general, as the IDS $N_0(E)$ is a monotone
increasing function, this result applies at all but a countable set of
energies.

In section 2, we recall the proof of Theorem 1.1. The proofs of Theorem 1.2,
and Corollaries 1.3-1.4, are given in section 3. We conclude with some
comments about the behavior of the density of states in section 4.
While preparing this letter,
we learned that Germinet and Klein \cite{[GK]} have proved a version of
(\ref{disordercont11}) for intervals $J$ away from zero. We thank F.\ 
Germinet
(private communication) for showing us the use of (\ref{traceid1}) that
improves our original estimates on $q_2$.

It is clear that there are various generalizations of our results. For 
example,
hypothesis (H1) can be weakened to allow unbounded random variables with the
first two moments bounded.


\section{Sketch of the Proof of Theorem 1.1.}

For completeness, let us sketch the proof of Theorem 1.1 that appears in
\cite{[CHK]}. We assume hypotheses (H1)-(H2) and condition (\ref{cont1}) on 
the
IDS $N_0(E)$ for the background operator $H_0$. Let $\Delta \subset I$ be a
sufficiently small closed interval, and let $ \tilde{\Delta} \supset \Delta$ 
be
a bounded interval with $|\tilde\Delta| = \mathcal{O} ( | \Delta |^\alpha 
)$,
for some $\alpha \in (0 , 1)$. First, one proves that (\ref{cont1}) implies
that for all $\Lambda$ sufficiently large, depending on $\tilde{\Delta}$, 
there
exists a finite constant $C_1(I, d) > 0$ so that
\beq
\label{unperttrace1}
Tr E_0^\Lambda ( \tilde{\Delta}) \leq C_1 ( I, d) |\tilde{\Delta}|^{q_1}
| \Lambda|.
\eeq

Next, we consider the local spectral projector $E_\Lambda ( \Delta)$ for
$H_\Lambda$ and write
\beq
\label{decomp1}
Tr E_\Lambda ( \Delta) = Tr
E_\Lambda ( \Delta) E_0^\Lambda ( \tilde{\Delta}) + Tr E_\Lambda ( \Delta )
E_0^\Lambda ( \tilde{\Lambda}^c),
\eeq
where $\tilde{\Delta}^c \equiv \R
\backslash \tilde{\Delta}$. The first term on the right in (\ref{decomp1}) 
is
easily seen to be bounded by
\beq
\label{unperttrace2}
Tr E_\Lambda ( \Delta)
E_0^\Lambda ( \tilde{\Delta}) \leq  Tr E_0^\Lambda(\tilde{\Delta}) \leq  C_1
(I, d) |\Delta|^{\alpha q_1} | \Lambda| ,
\eeq
and is already of order
$|\Delta|^q$, for any $q \le \alpha q_1$.

The second term on the right of (\ref{decomp1}) is estimated in second-order
perturbation theory. Let $E \in \Delta$ be the center of the interval 
$\Delta$,
and write \bea \label{unperttrace3} Tr E_\Lambda ( \Delta ) E_0^\Lambda
(\tilde{\Lambda}^c) & =& Tr E_\Lambda (\Delta) ( H_\Lambda - E )
        E_0^\Lambda (\tilde{\Lambda}^c) (H_0^\Lambda - E)^{-1} \nonumber \\
     & & - \lambda Tr E_\Lambda (\Delta) V_\Lambda E_0^\Lambda
(\tilde{\Lambda}^c)
     ( H_0^\Lambda - E )^{-1} \nonumber \\
      &=& (i) + (ii) .
\eea Since the distance from $\tilde{\Delta}^c$ to $E$ is of order
$|\Delta|^\alpha$, we easily see that term $(i)$ of (\ref{unperttrace3}) is
bounded as \beq \label{unperttrace4} |(i)| \leq |\Delta|^{1 - \alpha} Tr
E_\Lambda (\Delta ), \eeq so that as $0 < \alpha < 1$ and $|\Delta| < 1$, we
can move this term to the left in (\ref{decomp1}). Continuing with $(ii)$, 
we
repeat the calculation in (\ref{unperttrace3}), now to the left of
$E_\Lambda(\Delta)$, and obtain \bea \label{unperttrace5} (ii) &=& - \lambda 
Tr
(H_\Lambda - E) E_\Lambda (\Delta) V_\Lambda E_0^\Lambda (\tilde{\Lambda}^c)
     ( H_0^\Lambda - E )^{-2} \nonumber \\
   & & + \lambda^2 Tr V_\Lambda E_\Lambda (\Delta)V_\Lambda E_0^\Lambda
(\tilde{\Lambda}^c)( H_0^\Lambda - E )^{-2} \nonumber \\
&=& (iii) + (iv) .
   \eea
Term $(iii)$ is estimated as in (\ref{unperttrace4}) and we obtain \beq
\label{unperttrace6} | (iii)| \leq \lambda | \Delta|^{1 - 2 \alpha} \|
\tilde{V}_\Lambda \| Tr E_\Lambda (\Delta) , \eeq where $\tilde{V}_\lambda$ 
is
the potential obtained by replacing $\omega_j$ by the maximal value of
$|\omega_j|$. Term $(iv)$ in (\ref{unperttrace5}) can be bounded above by
\beq
\label{unperttrace7}
| (iv) | \leq \lambda^2 | \Delta|^{-2 \alpha} Tr V_\Lambda
E_\Lambda (\Delta) V_\Lambda .
\eeq

Taking the expectation and replacing $V_\lambda^2$ by the upper bound
$\tilde{V}_\Lambda^2$, we find that we must estimate \beq 
\label{unperttrace8}
\E \{ Tr ( \tilde{V}_\Lambda^2 E_\Lambda ( \Delta) ) \}. \eeq This is done
using estimates on the spectral shift function comparing the two local
Hamiltonians with one random variable fixed respectively at its maximum and
minimum values. For the lattice case, this is a rank one perturbation, so 
the
corresponding spectral shift is bounded by one, the rank of the 
perturbation.
For the continuous case, the perturbation is no longer finite rank, but we 
may
use the local $L^p$-estimate on the spectral shift function proved in
\cite{[CHN]}. In either case we obtain
\beq
\label{unperttrace9}
\E \{ Tr (
\tilde{V}_\Lambda^2 E_\Lambda ( \Delta) ) \} \leq C_4 (I,q^*,u) \lambda^{-1} 
|
\Delta|^{q^*} | \Lambda|,
\eeq
where the exponent $q^*$ in (\ref{unperttrace9}) is i)
$q^* = 1$ in the lattice case, ii) $0 < q^* < 1$ in the continuum.

As a consequence, term $(iv)$ in (\ref{unperttrace5}) can be bounded by \beq
\label{unperttrace10} \E \{ | (iv) | \} \leq \lambda
| \Delta |^{ q^* - 2 \alpha}
C_1 (I,d) C_4(I,q^*,u) | \Lambda|. \eeq Putting together 
(\ref{unperttrace2}),
(\ref{unperttrace4}), (\ref{unperttrace6}), and (\ref{unperttrace10}), we
obtain
\bea
\label{unperttrace11}
\lefteqn{ \{ 1 - | \Delta|^{1-\alpha} -
\lambda | \Delta |^{1 - 2 \alpha} \|
\tilde{V}_\Lambda \| \} \E \{ Tr E_\Lambda ( \Delta ) \} } \nonumber \\
& \leq & ( \lambda | \Delta |^{q^* - 2 \alpha} C_1(u,d) C_4(q^*,u) + C_2 |
\Delta|^{\alpha q_1} ) | \Lambda |.
\eea
By choosing the optimal $\alpha < 1/2$, it
is clear from this expression that Theorem 1.1 holds with $0 < q \leq q_1
q^* / ( q_1 + 2)$. $\Box$


\section{Proof of Theorem 1.2.}

The almost-sure existence of the IDS for random Hamiltonians of the type
considered here is well-known and we refer the reader to \cite{[CL], 
[Kirsch1],
[PF]}. The IDS $N_\lambda (E)$ is given in terms of the spectral projector
$P_\lambda (E)$ associated with $H_\omega (\lambda)$ and the interval
$(-\infty, E] \subset \R$. For the lattice case, with Hilbert space $\ell^2 
(
\Z^d)$, the IDS $N_\lambda (E)$
is given by
\bea \label{defnids1} N_\lambda (E) &=& \E
\{ Tr \delta_{0} P_\lambda (E) \delta_{0} \} \nonumber \\
   &=& \E \{ \langle 0 | P_\lambda (E) | 0 \rangle \} ,
\eea
where $\delta_0$ is the function supported at $0$ and $|x \rangle$ is the
state at site $x \in \Z^d$.
For the continuous case on $\R^d$, the IDS $N_0(E)$
is given by
\beq \label{defnids2} N_\lambda (E) \equiv \E \{ Tr \chi_0 P_\lambda (E)
\chi_0 \} ,
\eeq
with $\chi_0$ the characteristic function on the unit cube in
$\R^d$. {\it To unify the notation, we will write $\chi_0$ for the
characteristic function on the unit cube as in (\ref{defnids2}) in the
continuous case, or for the projector $\delta_{0}$ as in (\ref{defnids1}) in
the lattice case.}

\noindent {\bf Proof of Theorem 1.2.} Fix $\lambda,\lambda' \in J$ and $E 
\in
I$. Choose $g \in C^4 (\R)$, depending on $E$, $\lambda$, and $\lambda'$, 
with
$0 \leq g \leq 1$ and \beq \label{cutoff1} g(s) = \left\{
\begin{array}{ll}
              1 & s \leq E \\
              0 & s \geq E + | \lambda - \lambda' |^\alpha ,
          \end{array}
  \right.
\eeq where $0 < \alpha \leq 1$ will be determined. The choice of $g \in C^4$
obeying \eqref{cutoff1} is basically arbitrary, however we require that
\beq
\label{deriv1}
\| g^{(j)} \|_\infty \leq C | \lambda - \lambda' |^{- j \alpha}
\; , \quad (j=1,2,4) \; ,
\eeq
with some constant independent of $E$,
$\lambda$, and $\lambda'$ (this can be done).

We have \bea \label{strategy1} N_\lambda (E) - N_{\lambda'} (E)  &=& 
N_\lambda
    (E) - \E \left \{ Tr \chi_0 [g (H_\lambda)]^2 \chi_0 \right \} \nonumber 
\\
    & &+   \E \left \{ Tr \chi_0 [g(H_{\lambda'} )]^2
    \chi_0 \right \}-
    N_{\lambda'} (E) \nonumber \\
     & & +  \E \left \{  Tr \chi_0 ( [g(H_\lambda)]^2 - [g(H_{\lambda'})]^2 
)
     \chi_0 \right \}
      .
\eea
The monotonicity of $N_\lambda (E)$ with respect to energy and the
properties of $g$ imply that
\beq
\label{idsbound1}
\E \left \{ Tr \chi_0 [g
(H_\lambda)]^2 \chi_0 \right \} \chi_0 \leq N_\lambda (E + |\lambda -
\lambda'|^\alpha ).
\eeq
It follows from Theorem 1.1 that
\bea \label{est1} \E
\left \{ Tr \chi_0 [g (H_\lambda)]^2 \chi_0 \right \} - N_\lambda (E) & \leq 
&
N_\lambda (E + |\lambda - \lambda'|^\alpha ) - N_\lambda (E)
   \nonumber \\
&\leq & C(q_1,I,J) | \lambda - \lambda'|^{\alpha q}    ,
\eea
for any $0 < q \leq q_1 q^*  / (q_1+2)$,
and an identical estimate holds for the second term on the
right in (\ref{strategy1}).

It remains to estimate the last term on the right in (\ref{strategy1}). 
Using
the identity $2(A^2 - B^2) =  \{ A(A-B) +(A-B)A \} + \{B (A-B) + (A-B)B\}$, 
we
can write the last term in (\ref{strategy1}) as
\begin{multline}\label{trace1}
  \E \left \{ Tr \chi_0 ( [g(H_\lambda)]^2 - [g(H_{\lambda'})]^2 )
\chi_0 \right \} \\
= \E \left \{ Tr \chi_0  g(H_\lambda)  ( g(H_\lambda) -
g(H_{\lambda'}) ) \chi_0 \right \}\\
+ \E \left \{ Tr \chi_0  g(H_{\lambda'})  ( g(H_\lambda) - g(H_{\lambda'}) )
\chi_0 \right \}\; ,
\end{multline}
where, to reduce the number of terms, we have made use of the following
identity: If $A_\omega$ and $B_\omega$ are $\Z^d$-ergodic operators such 
that
$\chi_0 A_\omega B_\omega \chi_0$ is trace class, then we have \beq
\label{traceid1} \E \left \{ Tr \chi_0 A_\omega B_\omega \chi_0 \right  \}  
=
\E \left \{ Tr \chi_0 B_\omega A_\omega \chi_0 \right \} . \eeq (We use this
identity in a more crucial way below.) We note that the trace norm $\| 
\chi_0
g(H_\lambda) \|_1$ is bounded uniformly in $\lambda \in J$ as well as in the
random couplings $\omega_j$. In the continuum, we are using here that
$H_\lambda$ is bounded from below.

We express the difference $( g(H_\lambda) - g(H_{\lambda'}) )$ in terms of 
the
resolvents using the Helffer-Sj\"ostrand formula, which we now recall (see, 
for
example, \cite{[Davies]} for details). Given $f \in C_0^{k+1} (\R)$, we 
denote
by $\tilde{f}_k$ an almost analytic extension of $f$ of order $k$, which is 
a
function $\tilde{f}_k$ defined in a complex neighborhood of the support of 
$f$
having the property that $\tilde{f}_k (x + i 0) = f(x)$ and that \beq
\label{hsj1} |
\partial_{\bar z}\tilde{f}_k (x+ iy) | \sim |f^{(k+1)} (x) | |y|^k , 
~~\mbox{as} ~~ |y|
\rightarrow 0, \eeq where $\partial_{\bar z} = \partial_x + i \partial_y$. 
For
the construction of such a function, which is not unique, we refer to
\cite{[Davies]}. Let $R_\lambda (z) = (H_\lambda - z)^{-1}$ denote the
resolvent of $H_\lambda$. For functions $g$ as in (\ref{cutoff1}), the
functional calculus gives \beq \label{hsj2} g(H_\lambda) -g(H_{\lambda'}) =
\frac{(\lambda - \lambda') }{\pi} \int_{\C }
\partial_{\bar{z}} \tilde{g} (z) R_\lambda (z) V_\omega R_{\lambda'} (z) 
d^2z,
\eeq with $\tilde g$ an extension of order $3$ (recall that $g \in C^4$).

Let us estimate the first term on the right in (\ref{trace1}). The estimate 
for
the second term is similar. We substitute the Helffer-Sj\"ostrand formula
(\ref{hsj2}) and find \begin{multline} \label{trace6} Tr \chi_0 g 
(H_\lambda) (
g(H_\lambda) - g(H_{\lambda'}) ) \chi_0 \\ = \frac{(\lambda - \lambda') 
}{\pi}
\int_{\C }
\partial_{\overline{z}} \tilde{g} (z) Tr \chi_0 g (H_\lambda) R_\lambda (z)
V_\omega R_{\lambda'} (z) \chi_0 d^2z . \end{multline} Using the second
resolvent identity, we rewrite the operator involving resolvents as \bea
\label{improve1} R_\lambda (z) V_\omega R_{\lambda'} (z) &=& R_\lambda (z)
V_\omega (
R_{\lambda'} (z) - R_\lambda (z) )  + R_\lambda (z) V_\omega R_\lambda (z) 
\nonumber \\
&=& (\lambda' - \lambda ) R_\lambda (z) V_\omega R_{\lambda'} (z) V_\omega
R_{\lambda} (z) \nonumber \\
&& + R_\lambda (z) V_\omega R_\lambda (z) \nonumber \\
&=& (i) + (ii) . \eea

The integral in (\ref{trace6}) involving the first term $(i)$ in
(\ref{improve1}) is estimated as follows. The resolvents are bounded by $| 
\Im
z|^{-1}$ as $| \Im z| \rightarrow 0$, but this divergence is canceled by the
estimate \eqref{hsj1} for $\partial_{\bar z} \tilde g$ (since we take an
extension of order $4$). Recalling the estimate \eqref{deriv1} on the
derivatives of $g$ and noting $| \mbox{supp} ~g' | \sim \delta_1^{-1}$, with
$\delta_1 = | \lambda - \lambda'|^\alpha$, we find that \beq \int_{\C}
\abs{\partial_{\bar z} \tilde g(z)} |\Im z|^{-3} d^2z\le C|\lambda
-\lambda'|^{-3\alpha} ,\eeq since we obtain a factor of $\delta_1^{-4}$ from
the derivatives and a factor of $\delta_1$ from the size of the domain of
integration. Consequently, we find \begin{multline} \label{improve2}\E \left 
\{
\left| \frac{(\lambda - \lambda')^2 }{\pi} \int_{\C }
\partial_{\overline{z}} \tilde{g} (z) Tr \chi_0 g(H_\lambda) R_\lambda (z)
V_\omega R_{\lambda'} (z) V_\omega R_{\lambda}
(z)  \right| \right \} \\
\leq  C \E \left \{ \| \chi_0 g(H_\lambda) \|_1 \| V_\omega \|_\infty^2 
\right
\} | \lambda - \lambda' |^{2 - 3 \alpha} \leq C_0 |\lambda - \lambda'|^{2 - 
3
\alpha}.
\end{multline}

To evaluate the integral involving the second term $(ii)$ of 
(\ref{improve1}),
we apply \eqref{traceid1} to the operator integrand in $(ii)$ of
(\ref{improve1}). Inserting this into (\ref{trace6}), we obtain for the
integrand \beq \label{hsj4} \E \left \{ Tr \chi_0 g(H_\lambda) ( R_\lambda 
(z)
V_\omega R_{\lambda} (z) \chi_0 ) \right \}  = \E \left \{ Tr \chi_0 g
(H_\lambda) R_\lambda (z)^2 V_\omega \chi_0 \right \}. \eeq The integral
becomes \beq \label{hsj5} \int_{\C }
\partial_{\overline{z}} \tilde{g} (z) Tr \chi_0 g (H_\lambda) R_\lambda 
(z)^2
V_\omega \chi_0 d^2 z = - Tr \chi_0 g' (H_\lambda ) V_\omega \chi_0 . \eeq 
As a
result, we obtain the following estimate for the term involving $(ii)$ \beq
\label{improve3} \left | \E \{ Tr \chi_0 g' (H_\lambda ) V_\omega \chi_0 \}
\right | \leq C_1 | \lambda - \lambda'|^{-\alpha} . \eeq

Combining the estimates (\ref{est1}), (\ref{improve2}), and 
(\ref{improve3}),
we obtain the upper bound for the right side of (\ref{strategy1}),
\begin{multline} \label{result1}
| N_\lambda (E) - N_{\lambda'} (E)| \\
\leq 2 C(q_1,I) | \lambda - \lambda'|^{\alpha q} + C_0 | \lambda -
\lambda'|^{2-3 \alpha}+ C_1 | \lambda - \lambda'|^{1 - \alpha } .
\end{multline} Comparing the exponents of $|\lambda - \lambda'|$ in
(\ref{result1}), we can take $0 < \alpha < 1$
so that $\alpha q = 2 - 3 \alpha$,
giving the exponent $2 q / (q + 3)$. $\Box$

\vspace{.1in} The proof of Corollary 1.3 follows simply by taking $\lambda' 
=
0$. The continuity result of Corollary 1.4 is proved as follows.

\vspace{.1in} \noindent {\bf Proof of Corollary 1.4.} It suffices to note 
that
the proof of Theorem 1.1 in \cite{[CHK]} can be extended to prove that if 
$N_0
(E)$ continuous at $E$, then so is $N_\lambda (E)$. To see this, fix $E \in 
\R$
at which $N_0$ is continuous. Following the argument of \cite{[CHK]}, we see
that the finite-volume estimate (\ref{unperttrace1}) becomes the following. 
For
any $\epsilon>0$, there exists $\eta>0$ such that for
$\tilde\Delta=[E-\eta,E+\eta]$, and all $\Lambda$ sufficiently large, one 
has
\begin{equation}
\label{unperttrace21} Tr E_0^\Lambda ( \tilde{\Delta}) \leq \epsilon| 
\Lambda|.
\end{equation}
Without loss, we assume that $\eta < \epsilon$, since the left side of
\eqref{unperttrace21} is non-increasing in $\eta$.
Choose a closed subinterval
$\Delta=[E-\eta^p,E+\eta^p]$, with $p
>1$. Following the argument in section 2 with this choice of $\tilde\Delta$
and $\Delta$, the estimates (\ref{unperttrace4}), (\ref{unperttrace6}), and
(\ref{unperttrace10}) now have the form:
\begin{gather}
\label{unperttrace31}
  |(i)| \leq \eta^{p-1} Tr E_\Lambda (\Delta ),\\
  \label{unperttrace41}
  | (iii)|\leq C_0 \lambda \eta^{p-2}Tr E_\Lambda (\Delta)
  \\\label{unperttrace51}
  \E\{|(iv)|\}\leq C_1 \lambda\eta^{pq^*-2}|\Lambda|,
\end{gather}
where the constants $C_0$ and $C_1$ are independent of $\Lambda, \eta$, and
$\epsilon$ and the exponent $q^*$ appears in (\ref{unperttrace9}).
These imply that (\ref{unperttrace11}) has the form
\begin{equation}
  \{ 1 -\eta^{p-1} - C_0 \lambda \eta^{p-2}\}\E\{Tr E_\Lambda(\Delta)\}  
\leq
C (\lambda\eta^{pq^*-2}+\epsilon) | \Lambda |.
\end{equation}
If we pick $p > 3/q^*$, then for sufficiently small $\epsilon$, we get for 
all
$\Lambda$ sufficiently large
\begin{equation}
  \E\{Tr E_\Lambda(\Delta)\}\leq \epsilon C |\Lambda |,
\end{equation}
for some finite constant $C > 0$
since $\eta < \epsilon$. This shows that the IDS $N_\lambda$ is
continuous at $E$.

To complete the proof of Corollary 1.4, we return to equations
(\ref{strategy1}) and (\ref{est1}). We use the continuity of $N_\lambda (E)$ 
to
control the first and the last terms on the right in (\ref{strategy1}). For
example, we need to estimate \beq \label{oneterm1} N_\lambda (E) - \E \left 
\{
Tr \chi_0 g (H_\lambda) \chi_0 \right \}. \eeq The monotonicity of 
$N_\lambda
(E)$ with respect to energy, and the properties of $g$, imply that \beq
\label{idsbound11} \E \left \{ Tr \chi_0 g (H_\lambda) \chi_0 \right \} \leq
N_\lambda (E + |\lambda - \lambda'|^\alpha ) . \eeq It follows from the
continuity and monotonicity in $E$ of $N_\lambda (E)$ that \beq 
\label{est11} 0
\leq \E \left \{ Tr \chi_0 g (H_\lambda) \chi_0 \right \} - N_\lambda (E) 
\leq
N_\lambda (E + |\lambda - \lambda'|^\alpha ) - N_\lambda (E)
\xrightarrow[]{\lambda'\rightarrow \lambda} 0. \eeq
The estimate for the middle
term of (\ref{strategy1}) remains the same. Consequently, we have that \beq
\label{result2} \lim_{\lambda' \rightarrow \lambda} N_{\lambda'} (E) =
N_\lambda (E), \eeq at any point $E$ of continuity of $N_0$, proving 
Corollary
1.4. $\Box$

\vspace{.1in} \noindent {\bf Remark:} This proof shows that in general one 
can
control the modulus of continuity for the IDS $N_\lambda (E)$ of the random
model using that of the free model.


\section{Additional Comments and Conjectures}

In certain situations, we are able to obtain
more information about the density of states (DOS) $\rho_\lambda (E)$
and its behavior as
$\lambda \rightarrow 0$. The DOS is the derivative of the IDS $N_\lambda
(E)$ with respect to energy. Since the spectral shift function is pointwise
bounded for the lattice model, it follows from \cite{[CHK]} that the D0S
is bounded except at possibly a countable set of energies.
In this case, the DOS is given by
\beq
\label{dos1}
\rho_\lambda (E) \equiv \frac{d N_\lambda }{d E } (E) = \lim_{\epsilon
\rightarrow 0} \E \{ \Im \langle 0 | (H_\lambda - E - i \epsilon)^{-1} | 0
\rangle \} .
\eeq
Let us suppose that the random variables $\omega_j$ are Gaussian with mean
zero. In this case, the almost-sure spectrum of $H_\omega (\lambda)$ is
$\R$, for $\lambda \neq 0$, and the spectrum of $H_0 = \Delta$ is $[- 2d,
2d]$. If $E \in \R \backslash [ -2d, 2d]$, the resolvent can be expanded in
a Neumann series,
\beq
\label{dos2}
R_\lambda (E + i \epsilon ) = \Sum_{k=0}^\infty R_0 (E + i \epsilon) \left[
- \lambda V_\omega R_0 ( E+ i \epsilon ) \right]^k.
\eeq
The matrix elements of the free resolvent decay exponentially by the
Combes-Thomas argument. Let $d_0(E)$ be the distance from the spectrum of
$H_0$ to $E$. We then have the bound,
\beq
\label{dos3}
| \langle x| R_0 (E) |y \rangle | \leq \frac{C_0}{d_0(E)}e^{- d_0(E) |x-y|
   / 2 }  .
\eeq
We take the expectation of the zero-zero
matrix element in (\ref{dos2}). We expand the potentials $V_\omega$ and use
the estimate (\ref{dos3}) to control the sum over sites. We
easily see that the power series converges absolutely
provided
\beq
\label{dos4}
\frac{|\lambda| \E \{ | \omega_0 | \} C_1(d)} { d_0(E)^{d+1}} < 1,
\eeq
where the constant $C_1(d)$ depends on $C_0$ in (\ref{dos3}) and the
dimension. For example, for all $\lambda < 1$, we have the convergent 
expansion
\beq
\label{dos5}
\rho_\lambda (E) = \lambda^2 \rho^{(2)} (E) + \Sum_{k=3}^\infty \lambda^k
\rho^{(k)} (E) ,
\eeq
for all $|E| > [ C_1(d) \E \{ | \omega_0 | \} ]^{1/ (d+1) } + 2d$.

This result, and the results on the IDS in this note,
are steps towards proving the general conjecture concerning the
regularity of the DOS.
In particular, under the hypotheses (H1)--(H2),
we expect that the IDS is Lipschitz continuous, that is, we have $q=1$ in
(\ref{chk1}), with a constant independent of $\lambda$.
Furthermore, if the unperturbed
operator $H_0$ has a Lipschitz continuous IDS, then we expect that
\beq
\label{dos6}
| \rho_\lambda (E) - \rho_0 (E) | \leq C_q | \lambda|^q,
\eeq
for some constant $0 < C_q < \infty$, independent of $\lambda$, and some $0
< q \leq 1$.
Finally, if the distribution function for the random variable $\omega_0$ is
sufficiently regular, we expect that the IDS is also regular.



\begin{thebibliography}{99}

\bibitem{[CL]} R. Carmona, J. Lacroix, {\it Spectral theory of random
    Schr{\"o}dinger operators}, Boston: Birkha{\"u}ser, 1990.

\bibitem{[CHK]} J.\ M.\ Combes, P.\ D.\ Hislop, F.\ Klopp:
H\"older continuity of the integrated density of states for some random
operators at all energies, {\it International Mathematics Research Notices}
{\bf 2003}, 179--209 (2002).

\bibitem{[CHN]} J.\ M.\ Combes, P.\ D.\ Hislop, S.\ Nakamura: The
  $L^p$-theory of the spectral shift function, the Wegner estimate,
  and the integrated density of states for some random operators, {\it
    Commun.\ Math.\ Phys.} {\bf 218}, 113--130 (2001).

\bibitem{[CHKN]} J.\ M.\ Combes, P.\ D.\ Hislop, F.\ Klopp, S.\ Nakamura:
The Wegner estimate and the integrated density of states for some random
operators, {\it Proc.\ Indian Acad.\ Sci. (Math.\ Sci.)} {\bf 112}, 31--53
(2002).

\bibitem{[CHT]} J.\ M.\ Combes, P.\ D.\ Hislop, A.\ Tip:
Band edge localization and the integrated density of states for acoustic and
electromagnetic waves in random media, {\it Ann.\ Inst.\ Henri Poincar\'e} 
{\bf
70}, 381--428 (1999).

\bibitem{[Davies]} E.\ B.\ Davies:
Spectral theory and differential operators.\ Cambridge Studies in Advanced
Mathematics, {\bf 42}.\ Cambridge University Press, Cambridge, 1995

\bibitem{[FK1]} A.\ Figotin, A.\ Klein:
Localization of classical waves I: acoustic models, {\it Commun.\ Math.\ 
Phys.}
{\bf 180}, 439--482 (1996).

\bibitem{[FK2]} A.\ Figotin, A.\ Klein:
Localization for classical waves II: electromagnetic waves, {\it Commun.\
Math.\ Phys.} {\bf 184}, 411--441 (1997).

\bibitem{[GK]} F.\ Germinet, A.\ Klein (in preparation).

\bibitem{[Kirsch1]} W.\ Kirsch: Random \Schr\ operators: A course, in
  {\it \Schr\ operators, Sonderborg DK 1988}, ed.\ H.\ Holden and A.\
  Jensen, Lecture Notes in Physics {\bf 345}, Berlin: Springer 1989.

\bibitem{[PF]} L.\ Pastur, A.\ Figotin:
Spectra of random and almost-periodic operators.\ Berlin: Springer-Verlag,
1992.

\bibitem{[S]} J.\ H.\ Schenker: H\"older equicontinuity of the density of
states at weak disorder, to appear in {\it Lett.\ Math.\ Phys.}

\bibitem{[Wegner]} F.\ Wegner: The density of states for disordered
  systems, {\it Zeit.\ Phy.\ B} {\bf 44}, 9--15 (1981).


\end{thebibliography}
\end{document}